\documentclass{ws-procs961x669}            
\begin{document}
\title{Charge Emission from Near-Extremal Charged Black Holes}

\author{Chiang-Mei Chen$^{*}$}

\address{Department of Physics, National Central University, Chungli 320, Taiwan\\
$^*$E-mail: cmchen@phy.ncu.edu.tw}

\author{Sang Pyo Kim$^{\S}$}

\address{Department of Physics, Kunsan National University, Kunsan 54150, Korea\\
and Institute of Theoretical Physics, Chinese Academy of Sciences, Beijing 100190, China\\
$^\S$E-mail: sangkim@kunsan.ac.kr}

\author{Jia-Rui Sun$^{\dag}$}

\address{School of Physics and Astronomy, Sun Yat-Sen University, Guangzhou 510275, China\\
$^\dag$E-mail: sunjiarui@sysu.edu.cn}

\begin{abstract}
Using the symmetry of the near-horizon geometry and applying quantum field theory of a complex scalar field, we study the spontaneous pair production of charged scalars from near-extremal rotating, electrically and/or magnetically charged black holes. Analytical expressions for pair production, vacuum persistence and absorption cross section are found, and the spectral distribution is given a thermal interpretation. The pair production in near-extremal black holes has a factorization into the Schwinger effect in AdS and Schwinger effect in Rindler space, measuring the deviational from extremality. The associated holographical correspondence is confirmed at the 2-point function level by comparing the absorption cross section ratio as well as the pair production rate both from the gravity and the conformal field theories. The production of monopoles is discussed.
\end{abstract}


\bodymatter

\section{Introduction}\label{aba:sec1}
Black holes, as a consequence of vacuum fluctuations, create all species of particle pairs near their horizons, half of which radiate to the spatial infinity, leading to the so-called Hawking radiation.~\cite{Hawking:1974rv, Hawking:1974sw} Schwinger mechanism has also been known for long that virtual pairs created from vacuum fluctuations could be physically separated by a strong external electric field and substantiated as real pairs leading to spontaneous pair production.~\cite{Sauter:1931ab, Sauter:1932ab, Schwinger:1951nm}
Charged black hole thus provide both strong gravitational and electric fields resulting in either the Hawking radiation and/or Schwinger pair production, which intermingles part of quantum gravity effects and strong field quantum electrodynamics (QED) effects. The emission of charges from charged black holes has been studied independently of the Hawking radiation~\cite{Zaumen:1974, Carter:1974yx, Damour:1974qv, Gibbons:1975kk, Damour:1976jd} and intensively studied since then both for nonextremal and extremal cases~\cite{Ternov:1986bf, Khriplovich:1999gm, Khriplovich:1998si, Kim:2004us} (for a review and references, see Refs.~\citenum{Khriplovich:2002qn} and~\citenum{Ruffini:2009hg}).

In this paper, we review the emission of scalar charges from near-extremal charged black holes from the view point of quantum field theory in near-horizon geometry. A near-extremal black hole has an extremely small Hawking temperature, and therefore its Hawking radiation is exponentially suppressed. On the other hand, the electric field on the horizon becomes strong enough to trigger the emission of charges dominantly  through the Schwinger mechanism. This study differs from those early works in the following aspect. A conventional wisdom was to calculate the tunneling probability of virtual particles in negative energy states of the Dirac sea into positive energy states, whose mass gap is modified by the charged black hole to allow finite tunneling probability.~\cite{Damour:1974qv, Gibbons:1975kk, Damour:1976jd, Ternov:1986bf, Khriplovich:1999gm, Khriplovich:1998si, Kim:2004us} In the phase-integral method the leading contribution to the pair production comes from the poles near the horizon.~\cite{Kim:2016dmm} In our approach, we make use of the local geometry near the horizon: the geometry of a near-extremal charged black hole has the structure of $\text{AdS}_2 \times \text{S}^2$ for a nonrotating one and it has the warped $\text{AdS}_3$ for a rotating one, while the geometry of a non-extremal charged black hole is $\text{Rindler}_2 \times \text{S}^2$ for a nonrotating one. Hence, the symmetry of near-horizon geometries allows the separation of quantum fields, which leads to analytical expressions for the in-vacuum and the out-vacuum and thereby the spectral distribution of spontaneously created pairs. The Schwinger mechanism and QED vacuum polarization in $\text{dS}_2$ and $\text{AdS}_2$ and their relation to black holes are discussed in Ref.~\citenum{Kim:2016dmm}.

The Schwinger mechanism and the Hawking radiation as quantum tunneling~\cite{Parikh:1999mf} are, however, intertwined for the spontaneous pair production occurring in charged black holes, whose dominant contribution comes from the near-horizon region. Here we review and study the scalar particle emission in the spacetime of the near-horizon region of the near-extremal charged black holes. The simplest model is the scalar production in Reissner-Nordstr{\"o}m (RN) black holes~\cite{Chen:2012zn} (see Ref.~\citenum{Chen:2014yfa} for spinor production). In this case the spacetime has an $\text{AdS}_2 \times \text{S}^2$ structure and the electric field is constant near the horizon. Using the symmetry of geometry, one can analytically solve the Klein-Gordon (KG) equation and give an exact expression for the production rate. The analysis can be generalized to Kerr-Newman (KN) black holes~\cite{Chen:2016caa} and also include a magnetic charge.~\cite{Chen:2017mnm} Then, the angular momentum deforms the near-horizon spacetime geometry to be a warped $\text{AdS}_3$, but the KG equation still can be separated and solved analytically. The pair production has a remarkable thermal interpretation based on the discussions in Refs.~\refcite{Cai:2014qba,Kim:2015kna,Kim:2015qma,Kim:2015wda} and~\refcite{Kim:2016dmm}. Moreover, the scalar production has a nice conformal field theory (CFT) dual picture supporting the KN/CFTs correspondence.~\cite{Chen:2011gz, Chen:2012np}

\section{Emission of Charges from near-extremal KN black holes}
The near-horizon geometry of a near-extremal dyonic KN black hole has the structure of a warped $\text{AdS}_3$ as in Ref.~\refcite{Chen:2017mnm}
\begin{eqnarray} \label{NHKN}
ds^2 &=& \Gamma(\theta) \left[ -(\rho^2 - B^2) d\tau^2 + \frac{d\rho^2}{\rho^2 - B^2} + d\theta^2 \right] + \gamma(\theta) (d\varphi + b \rho d\tau)^2, \label{new coordinates}
\\
A_{[1]} &=& - \frac{Q (r_0^2 - a^2 \cos^2\theta) - 2 P r_0 a \cos\theta}{\Gamma(\theta)} \rho d\tau
\nonumber\\
&&- \frac{Q r_0 a \sin^2\theta - P (r_0^2 + a^2) \cos\theta \pm P\Gamma(\theta)}{\Gamma(\theta)} d\varphi, \label{A1n}
\end{eqnarray}
where
\begin{eqnarray}
&& \Gamma(\theta) = r_0^2 + a^2 \cos^2\theta, \qquad \gamma(\theta) = \frac{(r_0^2 + a^2)^2 \sin^2\theta}{r^2_0 + a^2 \cos^2\theta},
\nonumber\\
&& b = \frac{2 a r_0}{r_0^2 + a^2}, \qquad r_0 = \sqrt{Q^2 + P^2 + a^2}.
\end{eqnarray}
Here, $a$ is the angular momentum parameter and $Q, P$ are the electric and magnetic charges of the original dyonic KN black holes, and $B$ measures a deviation from the extremal limit and acts as the new horizon $\rho_H = B$ in this geometry. The zero magnetic charge $(P = 0)$ corresponds to the KN black hole while the zero angular momentum $(a = 0)$ corresponds to the RN black hole with both electric and magnetic charges and the gauge potential for the electric charge and Dirac monopole in the near-horizon geometry
\begin{eqnarray}
A_{[1]} = -Q \rho d \tau + P \bigl(\cos \theta \mp 1 \bigr) d \varphi.
\end{eqnarray}

Associated to the black hole thermodynamics, the Hawking temperature, entropy, angular velocity, and chemical potentials ($\bar\Phi_\mathrm{H}$ is given from the Hodge dual of Maxwell field $d A_{[1]}$ in the original KN metric) are
\begin{eqnarray} \label{BHT}
&& T_\mathrm{H} = \frac{B}{2 \pi}, \qquad S_\mathrm{BH} = \pi (r_0^2 + a^2 + 2 B r_0), \qquad \Omega_\mathrm{H} = - \frac{2 a r_0 B}{r_0^2 + a^2},
\nonumber\\
&& \Phi_\mathrm{H} = \frac{Q (Q^2 + P^2) B}{r_0^2 + a^2}, \qquad \bar\Phi_\mathrm{H} = \frac{P (Q^2 + P^2) B}{r_0^2 + a^2}. \label{therm var}
\end{eqnarray}
The thermodynamical variables~(\ref{therm var}), as the metric~(\ref{new coordinates}) and the gauge field~(\ref{A1n}) do, contain the KN black hole, RN black hole as well as dyonic RN black hole with both electric and magnetic charges in the limit of $P = 0$, $a = P = 0$, and $a = 0$, respectively. The KG equation for scalar dyons can be exactly solved. An effective potential due to the electromagnetic and gravitational interactions induces tunneling processes for the pair production. By imposing a suitable boundary condition, one can obtain the production rate from the incoming and outgoing fluxes on the horizon and the asymptotic boundary of the near-horizon geometry.~\cite{Chen:2012zn} For instance, without an incoming flux at the asymptotic outer boundary, the relative ratio of the outgoing (transmitted) flux in the asymptotic region to the incoming (reflected) flux at the horizon counts the spontaneously produced particles while the ratio of the outgoing (incident) flux to the incoming flux at the horizon gives the vacuum persistence amplitude due to vacuum fluctuations. On the other hand, the group velocity for created fermions leads to the relative ratio of the outgoing flux in the asymptotic region to the outgoing flux at the horizon for the fermion production probability.

We now focus on production of bosonic particles. Following Ref.~\citenum{Kim:2003qp}, the flux conservation
\begin{equation}
|D_{\text{incident}}| = |D_{\text{reflected}}| + |D_{\text{transmitted}}|,
\end{equation}
is related to the Bogoliubov relation
\begin{equation}
|\mathcal{A}|^2 - |\mathcal{B}|^2 = 1,
\end{equation}
where the vacuum persistence amplitude $|\mathcal{A}|^2$ and the mean number of produced pairs $|\mathcal{B}|^2$ are given by the ratios of the flux components
\begin{equation} \label{Bogoliubov}
|\mathcal{A}|^2 \equiv \frac{|D_{\text{incident}}|}{|D_{\text{reflected}}|}, \qquad |\mathcal{B}|^2 \equiv \frac{|D_{\text{transmitted}}|}{|D_{\text{reflected}}|}.
\end{equation}
Moreover, from the viewpoint of scattering of an incident flux from the asymptotic boundary, we can define the absorption cross section ratio as
\begin{equation} \label{absorptioncrosssection}
\sigma_{\text{abs}} \equiv \frac{|D_{\text{transmitted}}|}{|D_{\text{incident}}|} = \frac{|\mathcal{B}|^2}{|\mathcal{A}|^2}.
\end{equation}
In comparison, the group velocity for fermions leads to the vacuum persistence amplitude and the mean number of produced fermion pairs
\begin{equation} \label{fer Bogoliubov}
|\mathcal{A}|^2 \equiv \frac{|D_{\text{reflected}}|}{|D_{\text{incident}}|}, \qquad |\mathcal{B}|^2 \equiv \frac{|D_{\text{transmitted}}|}{|D_{\text{incident}}|},
\end{equation}
where the fluxes $D_{\text{incident}}, D_{\text{reflected}}$ and $D_{\text{transmitted}}$ are computed from the spin-diagonal equations.

Using the following ansatz (hereafter parameters $m, q$, and $p$ are the mass, electric and magnetic charges of a scalar field)
\begin{equation} \label{ansatz}
\Phi(\tau, \rho, \theta, \varphi) = \mathrm{e}^{-i \omega \tau + i [n \mp (q P - p Q)] \varphi} R(\rho) S(\theta),
\end{equation}
a straightforward calculation leads to the Bogoliubov coefficients and the absorption cross section ratio as (the derivation in detail can be found in Ref.~\citenum{Chen:2017mnm})
\begin{eqnarray}
|\mathcal{A}|^2 &=& \frac{\cosh(\pi \kappa - \pi \mu) \cosh(\pi \tilde{\kappa} + \pi \mu)}{\cosh(\pi \kappa + \pi \mu) \cosh(\pi \tilde{\kappa} - \pi \mu)},
\\
|\mathcal{B}|^2 &=& \frac{\sinh(2 \pi \mu) \sinh(\pi \tilde{\kappa} - \pi \kappa)}{\cosh(\pi \kappa + \pi \mu) \cosh(\pi \tilde{\kappa} - \pi \mu)}, \label{Bscalar}
\\
\sigma_{\text{abs}} &=& \frac{|\mathcal{B}|^2}{|\mathcal{A}|^2} = \frac{\sinh(2 \pi \mu) \sinh(\pi \tilde{\kappa} - \pi \kappa)}{\cosh(\pi \kappa - \pi \mu) \cosh(\pi \tilde{\kappa} + \pi \mu)}, \label{Sscalar}
\end{eqnarray}
where three essential parameters are ($\lambda_l$ being a separation constant)
\begin{equation} \label{kappamu}
\tilde \kappa = \frac{\omega}{B}, \quad \kappa = \frac{(q Q \!+\! p P) (Q^2 \!+\! P^2) \!-\! 2 n a r_0}{r_0^2 + a^2}, \quad \mu = \sqrt{\kappa^2 - m^2 (r_0^2 + a^2) - \lambda_l - \frac14},
\end{equation}
in which $\mu^2$ is positive due to the BF bound violation in the $\text{AdS}_2$ spacetime.~\cite{Chen:2017mnm}

Following our previous studies,~\cite{Chen:2012zn, Chen:2016caa} the mean number of produced pairs~(\ref{Bscalar}) can be rewritten as
\begin{equation} \label{meannumberpairs}
\mathcal{N} = |\mathcal{B}|^2 = \left( \frac{\mathrm{e}^{- 2 \pi \kappa + 2 \pi \mu} - \mathrm{e}^{- 2 \pi \kappa - 2 \pi \mu}}{1 + \mathrm{e}^{- 2 \pi \kappa - 2 \pi \mu}} \right) \left( \frac{1 - \mathrm{e}^{- 2 \pi \tilde\kappa + 2 \pi \kappa}}{1 + \mathrm{e}^{- 2 \pi \tilde\kappa + 2 \pi \mu}} \right).
\end{equation}
Note that the mean number~(\ref{meannumberpairs}) has a similar form as those of charged scalars with $\lambda_l = l(l+1)$ for spherical harmonics in a near-extremal RN black hole~\cite{Chen:2012zn} and with the same quantum number in KN black hole~\cite{Chen:2016caa} since the near-horizon geometry has an $\text{AdS}_2 \times S^2$ and a warped $\text{AdS}_3$ for the near-extremal RN and KN black holes, respectively. According to Refs.~\citenum{Kim:2015kna} and~\citenum{Kim:2015qma}, one may introduce an effective temperature and its associated counterpart
\begin{equation} \label{effectivetemperature}
T_\mathrm{KN} = \frac{\bar{m}}{2 \pi \kappa - 2 \pi \mu} = T_U + \sqrt{T_U^2 + \frac{\mathcal{R}}{8 \pi^2}}, \quad \bar{T}_\mathrm{KN} = \frac{\bar{m}}{2 \pi \kappa + 2 \pi \mu} = T_U - \sqrt{T_U^2 + \frac{\mathcal{R}}{8 \pi^2}},
\end{equation}
where the effective mass $\bar{m}$ is
\begin{equation} \label{barm}
\bar{m} = \sqrt{m^2 - \frac{\lambda + 1/4}{2} \mathcal{R}},
\end{equation}
and the corresponding Unruh temperature $T_U$ and AdS curvature $\mathcal{R}$ are
\begin{equation} \label{unruh}
T_U = \frac{\kappa}{2 \pi \bar{m}(r_0^2 + a^2)} = \frac{(q Q + p P)(Q^2 + P^2) - 2 n a r_0}{2 \pi \bar{m}(r_0^2 + a^2)^2}, \quad \mathcal{R} = - \frac{2}{r_0^2 + a^2}.
\end{equation}
Note that an AdS space binds a pair and thus increases the effective mass in contrast to a dS space which separates a pair and reduces the effective mass.

Finally, we may introduce a thermal interpretation by factorizing the mean number~(\ref{meannumberpairs}) as
\begin{equation} \label{thermal}
\mathcal{N} = \mathrm{e}^{\frac{\bar{m}}{T_\mathrm{KN}}} \times  \underbrace{\left( \frac{\mathrm{e}^{-\frac{\bar{m}}{T_\mathrm{KN}}} - \mathrm{e}^{-\frac{\bar{m}}{\bar{T}_\mathrm{KN}}}}{1 + \mathrm{e}^{-\frac{\bar{m}}{\bar{T}_\mathrm{KN}}}}\right) }_{\text{Schwinger effect in AdS$_2$}} \times \underbrace{\left\{ \frac{\mathrm{e}^{-\frac{\bar{m}}{T_\mathrm{KN}}} \left( 1 - \mathrm{e}^{-\frac{\omega - q \Phi_\mathrm{H} -p \bar\Phi_\mathrm{H} - n \Omega_\mathrm{H}}{T_\mathrm{H}}} \right)}{1 + \mathrm{e}^{-\frac{\omega - q \Phi_\mathrm{H} - p \bar\Phi_\mathrm{H} - n \Omega_\mathrm{H}}{T_\mathrm{H}}} \mathrm{e}^{- \frac{\bar{m}}{T_\mathrm{KN}}}} \right\} }_{\text{Schwinger effect in Rindler$_2$}}.
\end{equation}
The first parenthesis is the Schwinger effect with the effective temperature $T_\mathrm{KN}$ in $\text{AdS}_2$~\cite{Cai:2014qba} and the second parenthesis is the Schwinger effect in the Rindler space,~\cite{Gabriel:1999yz} in which the Unruh temperature is given by the Hawking temperature and charges have the chemical potentials $\Phi_\mathrm{H}, \bar\Phi_\mathrm{H}$, and $\Omega_\mathrm{H}$ while the effective temperature for the Schwinger effect due to the electric field on the horizon is still given by $T_\mathrm{KN}$. The extremal KN black hole can be obtained by taking the limit $T_\mathrm{H}=0$ and $\tilde \kappa = 0$, in which the second parenthesis multiplied by the prefactor becomes unity. An interesting physics is the monopole production: upon the Dirac quantization $(e p) = 2 \pi$ of charges and a small electric coupling $\alpha = e^2 = 1/137$, the symmetry of the pair production of electric and magnetic charges from dyonic black holes is broken due to $e/p = \alpha/2 \pi$. Monopole production and related physics go beyond the scope of this paper and will be addressed elsewhere.

\section{KN/CFT duality}
According to the KN/CFTs duality,~\cite{Chen:2010ywa, Chen:2012np} the absorption cross section ratio of scalar field in Eq.~(\ref{Sscalar}) corresponds to that of its dual operator in the dual two-dimensional CFT with left- and right-hand sectors
\begin{equation} \label{CFTabs}
\sigma_\mathrm{abs} \sim \frac{T_\mathrm{L}^{2 h_\mathrm{L} - 1} T_\mathrm{R}^{2 h_\mathrm{R} - 1}}{|\Gamma(2 h_L)| |\Gamma(2 h_R)|} \sinh\left( \frac{{\tilde\omega}_\mathrm{L}}{2 T_\mathrm{L}} \!+\! \frac{{\tilde\omega}_\mathrm{R}}{2 T_\mathrm{R}} \right) \! \left| \Gamma\left( h_\mathrm{L} \!+\! i \frac{{\tilde\omega}_\mathrm{L}}{2 \pi T_\mathrm{L}} \right) \right|^2 \! \left| \Gamma\left( h_\mathrm{R} \!+\! i \frac{{\tilde\omega}_\mathrm{R}}{2 \pi T_\mathrm{R}} \right) \right|^2,
\end{equation}
where $T_\mathrm{L}, T_\mathrm{R}$ are the temperatures, $h_\mathrm{L}, h_\mathrm{R}$ are the conformal dimensions of the dual operator, ${\tilde\omega}_\mathrm{L} = \omega_\mathrm{L} - q_\mathrm{L} \Phi_\mathrm{L}$ and ${\tilde\omega}_\mathrm{R} = \omega_\mathrm{R} - q_\mathrm{R} \Phi_\mathrm{R}$ are the total excited energy in which $(q_\mathrm{L}, q_\mathrm{R})$ and $(\Phi_\mathrm{L}, \Phi_\mathrm{R})$ are respectively the charges and chemical potentials (both including the electric and the magnetic contributions for the dyonic KN black hole case) of the dual left and right-hand operators. The complex conformal dimensions $(h_\mathrm{L}, h_\mathrm{R})$ of the dual operator be read out from the asymptotic expansion of the bulk dyonic charged scalar field at the AdS boundary~\cite{Chen:2017mnm}
\begin{equation} \label{cfmdim}
h_\mathrm{L} = h_\mathrm{R} = \frac12 \pm i \mu.
\end{equation}
For dyonic KN black holes, there are in general three different pictures, namely $J$-, $Q$- and the $P$-pictures, in the dual CFTs descriptions. Here, we only show the result of $J$-picture (for the other two pictures, see Ref.~\refcite{Chen:2017mnm}).

In the $J$-picture, the left- and right-hand central charges of the dual CFT are determined by the angular momentum~\cite{Chen:2010ywa, Chen:2012np}
\begin{equation}
c_\mathrm{L}^J = c_\mathrm{R}^J = 12 J,
\end{equation}
and the associated left- and right-hand temperatures for the near-extremal dyonic KN black hole are
\begin{equation}
T_\mathrm{L}^J = \frac{r_0^2 + a^2}{4 \pi a r_0}, \qquad T_\mathrm{R}^J = \frac{B}{2 \pi a}.
\end{equation}
The CFT microscopic entropy from the Cardy formula
\begin{equation}
S_\mathrm{CFT} = \frac{\pi^2}3 (c_\mathrm{L}^J T_\mathrm{L}^J + c_\mathrm{R}^J T_\mathrm{R}^J) = \pi (r_0^2 + a^2 + 2 r_0 B),
\end{equation}
agrees with the macroscopic entropy~(\ref{BHT}) of the near-extremal KN black hole.

Besides, by matching the first law of black hole thermodynamics with that of the dual CFT, i.e., $\delta S_\mathrm{BH} = \delta S_\mathrm{CFT}$, the following relation holds
\begin{equation}\label{1stlaws}
\frac{\delta M - \Omega_\mathrm{H} \delta J - \Phi_\mathrm{H} \delta Q - \bar\Phi_\mathrm{H} \delta P}{T_\mathrm{H}} = \frac{\tilde{\omega}_\mathrm{L}}{T_\mathrm{L}} + \frac{\tilde{\omega}_\mathrm{R}}{T_\mathrm{R}},
\end{equation}
where the angular velocity and chemical potentials at $\rho = B$ are given in Eq.~(\ref{BHT}). To probe the rotation we need to turn off the charges of the probe scalar field and set $T_\mathrm{L} = T_\mathrm{L}^J$ and $T_\mathrm{R} = T_\mathrm{R}^J$, then for the dyonic KN black hole $\delta M = \omega, \; \delta J = n, \; \delta Q = 0, \; \delta P = 0$. Thus, we have
\begin{equation}
{\tilde \omega}_\mathrm{L}^J = n \quad \mathrm{and} \quad {\tilde \omega}_\mathrm{R}^J = \frac{\omega}{a} \quad \Rightarrow \quad \frac{{\tilde\omega}_\mathrm{L}^J}{2 T_\mathrm{L}^J} = - \pi \kappa \quad \mathrm{and} \quad \frac{{\tilde\omega}_\mathrm{R}^J}{2 T_\mathrm{R}^J} = \pi \tilde\kappa,
\end{equation}
where $q, p$ are set to zero. Consequently, the agreement between the absorption cross section ratio~(\ref{Sscalar}) of the scalar field (with $q = p = 0$) in the near-extremal dyonic KN black hole and that of its dual scalar operator in Eq.~(\ref{CFTabs}) is confirmed in the $J$-picture.

\section{Conclusion}
We have reviewed and studied the emission of charges from near-extremal dyonic KN black holes within the framework of the quantum field in the near-horizon geometry.~\cite{Chen:2017mnm} The symmetry of the near-horizon geometry of near-extremal black holes leads to analytic expressions for the solutions and therefrom the in-vacuum and the out-vacuum and pair production rate. This makes a drastic difference from the conventional wisdom which computes the tunneling probability of virtual particles from the Dirac sea in charged black holes.~\cite{Zaumen:1974, Carter:1974yx, Damour:1974qv, Gibbons:1975kk, Damour:1976jd, Ternov:1986bf, Khriplovich:1999gm, Khriplovich:1998si, Khriplovich:2002qn, Kim:2004us, Ruffini:2009hg} In fact, the phase-integral formulation or the Hamilton-Jacobi equation for spinless charges has simple poles near the horizon, whose residues determine the leading term of pair production. On the other hand, the quantum field theory in the near-horizon geometry exhibits a rich structure of the spectral distribution of produced particles. A caveat, however, is that the exact solutions in the near-horizon are not the solutions in the original global spacetime of black holes but carry the essential information of pair production that occurs the near horizon and that the back-reaction due to produced pairs and the induced current is not considered.

The near-horizon region contains a causal horizon and the electric field effect near the horizon dominantly gives rise to both the Hawking radiation and the Schwinger mechanism. By imposing the proper boundary condition on the exact solutions, the pair production, vacuum persistence amplitude and absorption cross section can be obtained from the relative ratios of the fluxes on the asymptotic boundary and horizon. This approach provides one with a systematic method for the pair production and vacuum polarization in a proper-time integral representation.~\cite{Kim:2016dmm, Cai:2014qba} The thermal interpretation of the pair production rate for the near-extremal dyonic KN black holes consists of the Schwinger effect in the AdS$_2$ space mainly due to the electromagnetic field of black holes and the Schwinger effect in the Rindler space due to the Hawking temperature. The dual CFTs descriptions of the Schwinger pair production of the near-extremal KN black hole~\cite{Chen:2016caa} can be generalized into the threefold dual CFTs pictures for the dyonic KN black hole which includes an additional magnetic charge. The third $P$-picture is associated with the dual gauge potential, a new ``magnetic hair'' of the dyonic KN black hole and a $U(1)$ fiber on the base manifold. Based on the threefold dyonic KN/CFTs duality, the dual CFTs descriptions of the absorption cross section ratios and the pair production rate of the dyonic charged scalar field can be confirmed in the $J$-, $Q$-, and $P$-pictures, respectively.

The near-extremal dyonic KN black hole is the most general model in the sense that the zero magnetic charge limit corresponds to the KN black hole and the zero angular momentum (nonrotating) limit recovers the RN black hole with both electric and magnetic charge. Further restricting to zero angular momentum and magnetic charge reduces to the RN black hole.  The formulae in this paper have such limits by replacing the separation parameter $\lambda_l$ by $l (l+1)$ for spherical harmonics for nonrotating black holes. The magnetic charge has the Dirac monopole for the black hole, and the quantization condition $e p = 2 \pi$ of electric charges and the fine structure constant $e^2 = \alpha$ leads to $e/p = \alpha / 2 \pi$, which implies the magnetic charge is larger by order of three than the electric charge. Then, the monopole production from a magnetic black hole is suppressed since the Unruh temperature is proportional to $p / P$ compared to $e / Q$ of an electric black hole. The physics related to production of electric charges and magnetic monopoles from dyonic black holes will be addressed in a future publication.

\section*{Acknowledgments}
The authors would like to thank Gary Gibbons and Remo Riffini for pointing out that Schwinger mechanism was studied independently before Hawking radiation was discovered and also to express thanks to Alexei Gaina for informing us many relevant works related to this review article.
The work of C.M.C. was were supported by the Ministry of Science and Technology of the R.O.C. under the grant MOST 107-2119-M-008-013. The work of S.P.K. was supported by the Basic Science Research Program through the National Research Foundation of Korea (NRF) funded by the Ministry of Education (NRF-2015R1D1A1A01060626). The work of J.R.S. was supported by the NSFC under Grant No.~11675272 and the Fundamental Research Funds for the Central Universities.

\bibliographystyle{ws-procs961x669}

\begin{thebibliography}{10}

\bibitem{Hawking:1974rv}
  S.~W.~Hawking,
 {\it Nature} {\bf 248}, 30 (1974).

\bibitem{Hawking:1974sw}
  S.~W.~Hawking,
{\it  Commun.\ Math.\ Phys.}\  {\bf 43}, 199 (1975)
  Erratum: [{\it Commun.\ Math.\ Phys.}\  {\bf 46}, 206 (1976)].

 \bibitem{Sauter:1931ab} F.~Sauter,	
{\it Z.\ Phys.}\ {\bf 69}, 742 (1931).

\bibitem{Sauter:1932ab} F.~Sauter,
{\it Z.\ Phys.}\ {\bf 73}, 547 (1932).

\bibitem{Schwinger:1951nm}
  J.~S.~Schwinger,
{\it  Phys.\ Rev.}\  {\bf 82}, 664 (1951).

\bibitem{Zaumen:1974}
  W.~T.~Zaumen,
{\it Nature} {\bf 247}, 531 (1974).

\bibitem{Carter:1974yx}
  B.~Carter,
{\it  Phys.\ Rev.\ Lett.}\  {\bf 33}, 558 (1974).

\bibitem{Damour:1974qv}
  T.~Damour and R.~Ruffini,
{\it  Phys.\ Rev.\ Lett.}\  {\bf 35}, 463 (1975).

\bibitem{Gibbons:1975kk}
  G.~W.~Gibbons,
{\it  Commun.\ Math.\ Phys.}\  {\bf 44}, 245 (1975).

\bibitem{Damour:1976jd}
  T.~Damour and R.~Ruffini,
{\it  Phys.\ Rev.\ D} {\bf 14}, 332 (1976).

\bibitem{Ternov:1986bf}
  I.~M.~Ternov, A.~B.~Gaina and G.~A.~Chizhov,
{\it  Sov.\ J.\ Nucl.\ Phys.}\  {\bf 44}, 343 (1986)
  [{\it Yad.\ Fiz.}\  {\bf 44}, 533 (1986)].

\bibitem{Khriplovich:1998si}
  I.~B.~Khriplovich,
 {\it J.\ Exp.\ Theor.\ Phys.}\  {\bf 88}, 845 (1999)
  [{\it Zh.\ Eksp.\ Teor.\ Fiz.}\  {\bf 115}, 1539 (1999)]  [gr-qc/9812060].

\bibitem{Khriplovich:1999gm}
  I.~B.~Khriplovich,
{\it  Phys.\ Rept.}\  {\bf 320}, 37 (1999).

 \bibitem{Kim:2004us}
  S.~P.~Kim and D.~N.~Page,
{\it  Nuovo Cim.\ B} {\bf 120}, 1193 (2005)
  [gr-qc/0401057].

\bibitem{Khriplovich:2002qn}
  I.~B.~Khriplovich,
{\it  Phys.\ Atom.\ Nucl.}\  {\bf 65}, 1259 (2002)
  [{\it Yad.\ Fiz.}\  {\bf 65}, 1292 (2002)].

\bibitem{Ruffini:2009hg}
  R.~Ruffini, G.~Vereshchagin and S.~S.~Xue,
{\it  Phys.\ Rept.}\  {\bf 487}, 1 (2010)
  [arXiv:0910.0974 [astro-ph.HE]].

\bibitem{Kim:2016dmm}
  S.~P.~Kim,
 {\it Int.\ J.\ Mod.\ Phys.\ D} {\bf 25}, 1645005 (2016)
  [arXiv:1602.05336 [hep-th]].

\bibitem{Parikh:1999mf}
  M.~K.~Parikh and F.~Wilczek,
{\it  Phys.\ Rev.\ Lett.}\  {\bf 85}, 5042 (2000)
  [hep-th/9907001].

\bibitem{Chen:2012zn}
  C.-M.~Chen, S.~P.~Kim, I.-C.~Lin, J.-R.~Sun and M.-F.~Wu,
{\it  Phys.\ Rev.\ D} {\bf 85}, 124041 (2012)
  [arXiv:1202.3224 [hep-th]].

\bibitem{Chen:2014yfa}
  C.-M.~Chen, J.-R.~Sun, F.-Y.~Tang and P.-Y.~Tsai,
{\it  Class.\ Quant.\ Grav.}\  {\bf 32}, 195003 (2015)
  [arXiv:1412.6876 [hep-th]].

\bibitem{Chen:2016caa}
  C.-M.~Chen, S.~P.~Kim, J.-R.~Sun and F.-Y.~Tang,
{\it  Phys.\ Rev.\ D} {\bf 95}, 044043 (2017)
  [arXiv:1607.02610 [hep-th]].

\bibitem{Chen:2017mnm}
  C.-M.~Chen, S.~P.~Kim, J.-R.~Sun and F.-Y.~Tang,
{\it  Phys.\ Lett.\ B} {\bf 781}, 129 (2018)
  [arXiv:1705.10629 [hep-th]].

\bibitem{Cai:2014qba}
  R.~G.~Cai and S.~P.~Kim,
{\it  JHEP} {\bf 1409}, 072 (2014)
  [arXiv:1407.4569 [hep-th]].

\bibitem{Kim:2015kna}
  S.~P.~Kim, H.~K.~Lee and Y.~Yoon,
  arXiv:1503.00218 [hep-th].

\bibitem{Kim:2015qma}
  S.~P.~Kim,
{\it  Int.\ J.\ Mod.\ Phys.\ A} {\bf 30}, 1545017 (2015)
  [arXiv:1506.03990 [hep-th]].

\bibitem{Kim:2015wda}
  S.~P.~Kim,
  doi:10.1142/$9789814759816_-0011$
  [arXiv:1509.05532 [hep-th]].

\bibitem{Chen:2011gz}
  C.-M.~Chen and J.-R.~Sun,
{\it  J.\ Phys.\ Conf.\ Ser.}\  {\bf 330}, 012009 (2011)  [arXiv:1106.4407 [hep-th]].

\bibitem{Chen:2012np}
  C.-M.~Chen and J.-R.~Sun,
{\it  Int.\ J.\ Mod.\ Phys.\ Conf.\ Ser.}\  {\bf 07}, 227 (2012)
  [arXiv:1201.4040 [hep-th]].

\bibitem{Kim:2003qp}
  S.~P.~Kim and D.~N.~Page,
{\it  Phys.\ Rev.\ D} {\bf 73}, 065020 (2006)
  [hep-th/0301132].

\bibitem{Gabriel:1999yz}
  Cl.~Gabriel and Ph.~Spindel,
{\it  Annals Phys.}\  {\bf 284}, 263 (2000)
  [gr-qc/9912016].

\bibitem{Chen:2010ywa}
  C.-M.~Chen, Y.-M.~Huang, J.-R.~Sun, M.-F.~Wu and S.-J.~Zou,
{\it  Phys.\ Rev.\ D} {\bf 82}, 066004 (2010)  [arXiv:1006.4097 [hep-th]].


\end{thebibliography}



\end{document}